\documentclass[aps,twocolumn,prb,unsortedaddress,showpacs]{revtex4}
\usepackage{amssymb}
\usepackage{graphicx}
\usepackage{color}
\usepackage{epsfig}

\begin{document}

\title{Lattice Dynamics and  Specific Heat of $\alpha$ - GeTe:
a theoretical and experimental study}

\author{R. Shaltaf}
\author{X. Gonze}
\affiliation{
European Theoretical Spectroscopic Facility, Unit PCPM\\
Universit\'{e} Catholique de Louvain,
B-1348 Louvain-la-Neuve, Belgium}

\author{M. Cardona}
\author{R. K. Kremer}
\email[Corresponding author:~E-mail~]{r.kremer@fkf.mpg.de}
\author{G. Siegle}
\affiliation{Max-Planck-Institut f{\"u}r Festk{\"o}rperforschung,
Heisenbergstr. 1, D-70569 Stuttgart, Germany}

\date{\today}

\begin{abstract}
We extend recent \textit{ab initio} calculations of the electronic  band structure and the phonon dispersion relations of
rhombohedral GeTe to calculations of the density of phonon states and the temperature dependent specific heat.
The results are compared with measurements of the specific heat. It is discovered that
the specific heat
depends on hole concentration, not only in the very low temperature region (Sommerfeld term)
but also at the maximum of  $C_p/T^3$  (around 16 K). To explain this phenomenon,
we have performed \textit{ab initio} lattice dynamical calculations for
GeTe rendered metallic through the presence of a heavy hole concentration
($p$ $\sim$ 2$\times$ 10$^{21}$ cm$^{-3}$). They account for the increase observed in the maximum of $C_p/T^3$.
\end{abstract}

 \maketitle



\section{Introduction}
GeTe is a material with 10 valence electrons per primitive unit cell (PC) and is thus related to the lead
chalcogenides PbX (X = S, Se, Te) and to the semimetals As, Sb, and Bi. Like PbX,
GeTe is a semiconductor but crystallizes in the rock salt structure  only at
temperatures above $\sim$ 700K.~\cite{Rabe1987a}
Below $\sim$ 650K, GeTe distorts through an extension of the cube diagonal into a rhombohedral
structure~\cite{Rabe1987b} similar to that of the
semimetals As, Sb and Bi, except that the two atoms per PC are not equal and hence its bonding
is polar. Compatible with the rhombohedral symmetry represented by a strain along [111],
there are two additional independent distortion parameters, the rhombohedral angle $\alpha$,
whose distortion from the 60$^{\rm o}$ of the cubic phase brings it to $\approx$58$^{\rm o}$,  and a
change in the Ge-Te sublattice separation $a_0 \tau \sqrt{3}$ with $\tau \sim 0.03$.
\cite{Rabe1987a,Goldak1966}

This distortion  converts GeTe  into a ferroelectric with
$T_c \sim$  650 K, given the polar nature of the bond.  Another interesting property
of GeTe is its native $p$-type doping, with hole concentrations typically higher
than 5$\times$ 10$^{19}$ cm$^{-3}$. This fact is attributed to the presence of Ge vacancies.\cite{Bevolo1976,Hein1964}

The metallic nature of GeTe with high hole concentration ($p >$ 8.3$\times$10$^{20}$)
leads to superconductivity at temperatures $T_c$ below $\sim$ 0.3 K.\cite{Hein1964}
Finegold~\cite{Finegold1964} examined the heat capacity of a GeTe
sample with $p \sim$ 1.1$\times$10$^{21}$ cm$^{-3}$  and $T_c \sim$ 0.25 K in the region around $T_c$
and observed the peak related to superconducting behavior: application
of a magnetic field of 500 Oe completely wiped out superconductivity and
the related peak, thus allowing him to determine, for his sample, a Sommerfeld
term $\gamma$ = 1.32 mJ/moleK$^2$. The specific heat measurements reported here were performed
from 4 to 250 K, and therefore do not address the superconducting behavior of GeTe.
They allow us, however, to determine $\gamma$  and the maximum in $C_p/T^3$ ( at $\sim$17 K)
which signals deviations from Debye's $T^3$ law. We find this maximum to be 5.5\%
higher than that obtained from the data of Bevolo \textit{et al.}~\cite{Bevolo1976}, a fact that
suggests a decrease in phonon frequencies induced by the presence of hole doping whose
concentration is, in our sample, $\sim$25 times higher than in that of Bevolo \textit{et al.}.

In order to test this hypothesis we have performed \textit{ab initio} calculations
of the lattice dynamics of $\alpha$-GeTe similar to those reported earlier
for PbS~\cite{Cardona2007}  and for
PbS, PbSe, and PbTe~\cite{Romero2008}. Comparison of theoretical -obtained without spin-orbit (SO) coupling-
and experimental values of   $C_p/T^3$  presented in Ref. [\onlinecite{Cardona2007}] suggested
that there may be a significant contribution ($\sim$ 20\%) of the SO
interaction to the value of $C_p/T^3$ at its maximum
(note that in the region of the maximum
$C_p \approx C_v$.  \cite{Mills1974})

We therefore repeated in  Ref. \onlinecite{Romero2008} lattice dynamical
calculations for PbS, PbSe and PbTe, this time with and without SO interaction.
We were able to prove that the lack of SO interaction, especially at the Pb ions,
was largely responsible for the discrepancy just mentioned. In the calculations reported
here for $\alpha$-GeTe we have checked the effect of SO interaction and,
not unexpectedly (heavy Pb ions are not present), we found it to be rather small ($\sim$1\%);
it can thus be neglected in the remaining of this work. We did find, however, for the
calculated  values of $C_p/T^3$ an increase in its maximum between an undoped sample
and one with $p\sim$ 2.1$\times$10$^{21}$ cm$^{-3}$ of  $\sim$6\%,  which largely
explains the difference found between our sample and that of Bevolo \textit{et al.} (3.5\%).
Our calculations of the lattice dispersion relations for $p\sim$ 2.1$\times$10$^{21}$ cm$^{-3}$ and
for undoped samples show indeed that the phonon frequencies decrease with doping.
This effect is very large around the $\Gamma$-point of the Brillouin zone (BZ) due to
the screening of the ionic charge by the holes. However, it amounts to a few
percent throughout most of the BZ.

As a by-product we have also calculated the phonon density of states (DOS) and
its projection on each of the constituent ions. We have also calculated the density
of two-phonon states which applies to optical spectroscopies involving two phonons:
second order Raman scattering and infrared absorption. It is hoped that the
availability of these DOS calculations will encourage such optical measurements
in this interesting material .

\section{Theoretical Details}

All the calculations have been performed using plane waves and
norm-conserving pseudopotentials, as implemented in the ABINIT code.~\cite%
{ABINIT, abinit05} The  dynamical properties have been
evaluated within the density-functional perturbation theory.~\cite%
{Baroni01,XGonze10337,XGonze10355}
We employed Hartwigsen-Goedecker-Hutter (HGH) pseudopotentials,~\cite{HGH}
generated including spin-orbit coupling, within the local density
approximation (LDA) adopting the Teter-Pade parameterization.\cite{XC}
Even though HGH pseudopotentials are known as being
relatively hard, the properties investigated in this work are well converged
when including a plane wave basis up to a kinetic energy cutoff
equal to 15 Ha. The Brillouin zone integration
was performed using special \textbf{k}-points sampled within the Monkhorst-Pack scheme%
.~\cite{Monkhorst-Pack} We found that a mesh of $12\times12\times12$ \textbf{k}-points
was required to describe well the structural and vibrational properties.
Further details on the theoretical calculations are
mostly given in Ref. \onlinecite{Romero2008}
except those for the heavily
$p$-doped material which were only performed without SO interaction.

\begin{table}[b]
\caption{Calculated structural parameters of GeTe. The lattice
parameter
$a_0$ (in $\rm{\AA} $ ), the angle $\protect\alpha$ (in deg), the deviation
from of the Ge sublattice from the 0.5 sublattice position $\protect\tau$,
and the volume $\Omega=(a^{3}_{0}/4)\sin^{2}\protect\alpha$ (in $\rm{\AA}^3$). }

\label{structural}
\begin{center}
\begin{ruledtabular}
\begin{tabular}{c c c c c c c c    }
& &$a_0$   & $\alpha$& $\tau$ & $\Omega$     \\
        \hline\hline&&&\\
with SO && 5.897  & 88.96  &  0.0237 & 51.26   \\
without SO && 5.894  & 88.96  &  0.0237 & 51.19   \\
\hline
Theory~\cite{Ciucivara} && 5.886 &89.24  & 0.0217& 50.96    \\
Exp ~\cite{Goldak1966} &&5.996 & 88.18  &  0.026 & 53.84 \\
Exp ~\cite{Onodera} &&5.98 & 88.35  &  0.0248 &53.31   \\

\end{tabular}
\end{ruledtabular}
\end{center}
\end{table}

To mimic the effect of Ge vacancies on the calculated force constants, and consequently on the  heat capacity,
the number of total electrons was reduced by a certain amount $\delta e$. The value of $\delta e$
was  chosen to be equivalent to the concentration of the free holes  $p$.  In all calculations of hole doped cases,
a background of homogeneous negative charge was  imposed to achieve charge neutrality.

Before calculating the dynamical matrix the structural parameters $a_0$ , $a$, and $\tau$ were optimized.
In Table~\ref{structural} we show the calculated structural parameters with and without SO coupling. The effect of the SO
interaction on the structural parameters is found to be negligible. Our results
are globally in good
agreement with previously reported \textit{ab initio} results.~\cite%
{Ciucivara}. Moreover the deviation between the calculated structural
parameters($a_{0}$,$\alpha $) and experiment is less than 2\%.
We performed the calculations using the  parameters obtained without spin-orbit coupling
for all cases including those of doped samples.

\section{Calculated Dispersion Relations and Phonon Density of States}

\begin{figure}[tbph]
\includegraphics[width=8.5cm,clip=true]{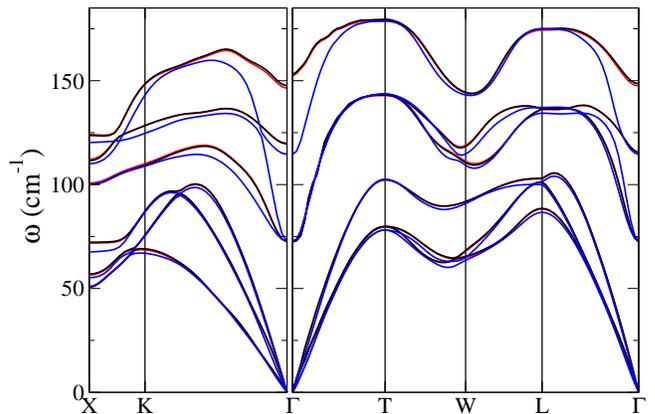}
\caption{(color online) Calculated phonon dispersion relations  of GeTe
along selected symmetry directions in the BZ. The (black) solid line denotes the
results for stoichiometric undoped GeTe, the (red) solid line denotes the
results including spin-orbit coupling and the (blue) solid line was obtained for doped GeTe
with $p=$ 2$\times$10$^{21}$holes/cm$^3$ without spin-orbit coupling.
} \label{fig1}
\end{figure}

In Fig. \ref{fig1} we show the phonon dispersion relations calculated for undoped GeTe
($p$ = 0) and for GeTe containing $p$ = 2.1$\times$10$^{21}$ holes/cm$^3$ . The undoped material, to
the best of our knowledge, has never been prepared. We calculate it as a reference point
to assess the effects of $p$-type doping on the dispersion relations. For $p$ = 0 we have
performed calculations without and with SO interaction. This interaction has a very weak
effect, only sizeable near the $\Gamma$-point of the BZ: the frequency of the LO phonons near $\Gamma$ is lowered by 1.3 cm$^{-1}$
when the SO interaction is included (see Table \ref{phonong}). This effect can be compared with that
found for PbTe~\cite{Romero2008} for the equivalent phonons, which lies around 5 cm$^{-1}$.
Assuming that this effect is proportional to the sum of SO splittings of the valence
electrons of cation and anion \cite{Romero2008} (for values of atomic SO
splittings see Ref \onlinecite{Herman1963}) we predict, from the value
of 5 cm$^{-1}$ calculated for PbTe, 1.4 cm$^{-1}$ for GeTe, compatible with the value of 1.3 cm$^{-1}$
obtained in the calculations. The softenings induced by SO coupling for other
phonons in the BZ are even smaller and thus will be henceforth neglected.

In Fig. \ref{fig1} we can also appreciate the effect of hole doping on the dispersion relations.
Without doping, the TO phonons are split by as much as 45 cm$^{-1}$ on account of the electrostatic
field induced by the large transverse ionic charges~\cite{Shaltaf2008}. This effect is a
non-analytic function of wave vector \textit{\textbf{k}} : for \textit{\textbf{k}} $\approx$ 0 the LO and TO frequencies depend on the direction
of $k$. The corresponding discontinuities disappear in the doped material, as the transverse charges are
screened by the free holes: like in the cases of As, Sb and Bi \cite{Diaz2007PRL,Serrano2008}
the optical phonons of the doped GeTe are split by the rhombohedral field into a singlet (LO) and
a doublet (TO), the splitting amounting to 42 cm$^{-1}$ , in agreement with measurements by
Raman spectroscopy.\cite{Andrikopoulos} and with our experimental results (see Table~\ref{phonong} and section IV-A).
The individual calculated frequencies  are, however, smaller than experimental results.
The differences between the absolute values of calculated and measured frequencies is probably due to the employed
local density and pseudopotential approximations, or, to temperature effects resulting from anharmonic terms which we do not include
in the present calculations.
Similar behavior has been noticed for the lead chalcogenides where the calculated LO frequency of PbTe is 28 cm$^{-1}$ lower than the measured one,
see Fig. 3 of Ref. \onlinecite{Romero2008}.

\begin{table}[b]
\caption{{Phonon frequencies at the zone center (in cm$^{-1}$) calculated for a
$\mathbf{q}$ vector $\parallel$ to trigonal axis.} }
\label{phonong}
\begin{center}
\begin{ruledtabular}
\begin{tabular}{ c c c }
             &      $E(TO)$  & $A_1(LO)$       \\
\hline\hline\\
$p=0$ (no spin-orbit)    &         73  &  153  \\
$p=0$ (with spin-orbit)  &         72  &  152  \\
\hline
$p=2.1\times10^{21}$ holes/cm$^3$   &         73  &  115  \\
Exp(present work)        & 88 & 123\\
Exp\cite{Andrikopoulos}&         80  &  122  \\

\end{tabular}
\end{ruledtabular}
\end{center}
\end{table}

We should mention at this point that similar softenings of the phonons of semiconductors upon
heavy doping have been investigated earlier, both experimentally and theoretically. Most recently
the case of boron doped diamond has been considered, in connection with the superconductivity observed
in this material.\cite{Boeri2006,Hoesch2007}
For a hole concentration similar to the one used here Boeri \textit{et al.} find for diamond a
softening of the optical frequency at $\Gamma$ of $\sim$20\%.
From Table~\ref{phonong}  we estimate a doping induced softening  of 15\% for the LO phonon
and 30\% for the TO phonons at $\Gamma$.
It has been pointed out that the strong phonon softening, in the case of diamond, is related to a
strong hole-phonon
interaction, probably also responsible for superconductivity.\cite{Cardona2006a}
Both, ferroelectricity and superconductivity, point to a strong carrier-phonon interaction in GeTe.
Phonon softenings upon doping with boron have also been observed
in Si.\cite{Pintschovius1982,Cerdeira1973}

We have calculated the phonon DOS  and its projections on the Ge and Te atoms using the same procedure as in Refs. \onlinecite{Cardona2007} and \onlinecite{Romero2008}.

\begin{figure}[tbph]
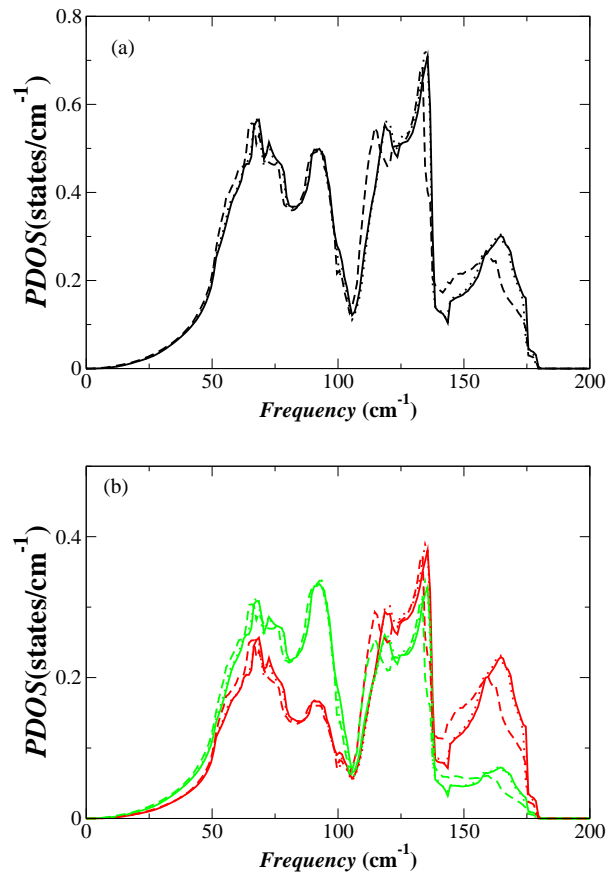

\begin{tabular}{c}
\epsfig{file=Fig-2a.eps,width=7.9cm,clip=true}\\\\
\epsfig{file=Fig-2b.eps,width=7.9cm,clip=true}\\
\end{tabular}
\caption{(color online) (a)
One-phonon DOS of $\alpha$-GeTe calculated for an intrinsic sample
(solid lines) without SO interaction, (dotted line) with SO interaction included,
and (dashed line) for a sample with $p$ = 2.1$\times$10$^{21}$ holes/cm$^3$.
(b) Projection of the DOS curves of Fig.2a on Ge (red) and on Te (green) atoms.}
\label{fig2}
\end{figure}

The one-phonon DOS is  shown in Fig.\ref{fig2}(a) for three different cases: undoped GeTe without SO interaction and
with SO interaction included (note that the difference is hardly visible and, as already mentioned above, can be neglected)
and GeTe with   $p$ = 2.1$\times$10$^{21}$ holes/cm$^3$  (dashed line, no SO interaction). The presence of holes at the top
of the valence band
lowers all of the three bands in Fig. \ref{fig2}(a), as corresponds to the softening of the dispersion relations
in Fig. \ref{fig1}. Figure \ref{fig2}(b) displays the three DOS of  Fig. \ref{fig2}(a) projected on the  two constituent atoms of $\alpha$-GeTe.
In Figure \ref{fig2} three bands are observed. They correspond to the acoustic phonons (from 0 to 110 cm$^{-1}$), the TO, and the LO phonons, respectively.
Whereas in PbX (X=S, Se, Te) \cite{Romero2008} the TA band is almost exclusively Pb-like, Fig. \ref{fig2}(b) shows that for GeTe this
band contains a strong mixture of Te and Ge vibrations ($\sim$60-40\%)  whereby the heavier ion dominates, as expected for acoustic modes.
The intermediate (TO) band contains a nearly equal mixture of vibrations of both atoms, with a slight predominance of the
lighter one ($\sim$ 90-10 \%).
The LO-band is clearly dominated by Ge vibrations ($\sim$30-70\%), as expected. All bands are lowered
in the heavily $p$-doped material with
respect to the undoped ones, as expected from the dispersion relations in Fig. \ref{fig1}. This lowering
results in an increase in the maximum of  $C_{v,p}$ vs. $T$ found at $\sim$16K, as will be discussed in Sect. \ref{secexp}.

\section{EXPERIMENTAL PROCEDURE AND RESULTS}\label{secexp}

\subsection{Samples and Sample Characterization}

Most of the measurements reported here were performed on wafers of a coarse polycrystalline ingot piece grown by
the Bridgman technique in a quartz ampoule.\cite{Schoenherr1980} The ingot diameter was about 10 mm. We cut from it several wafers
perpendicular to the growth axis, about 0.8 mm in thickness. The wafer measured by us will be labeled GeTe-CG.
We also measured the heat capacity of commercial samples (99.999\% purity, Alfa Aesar, Karlsruhe, Germany) consisting of small
crystalline chips, about 1 mm in size.
Several chips weighing a total of  $\sim$23 mg were measured (they are labeled as Alfa-GeTe below). The results are
compared with the data published by Bevolo \textit{et al.}~\cite{Bevolo1976}.

The wafers were characterized by Hall effect measurements using the Van der Pauw
technique.\cite{vanderPauw1958}.
No such measurements
were possible for the granular Alfa Aesar
sample. In this case, the hole concentration $p$ was estimated from the linear-in-$T$
term of the specific heat (Sommerfeld term, equal to $\gamma T$) by comparing it with that
reported in Ref. \onlinecite{Bevolo1976} for $p$ = 8$\times$10$^{19}$ holes/cm$^3$. For this purpose we used the relation~\cite{Ashcroft}:

\begin{equation}
\gamma \propto m^* p^{1/3}
\label{eq1}
\end{equation}

where $m^*$ is the effective band mass of the holes.
We also estimated $p$ in the CG2-GeTe sample from the plasma minimum in the ir reflection
spectrum by comparison with similar data in Ref. \onlinecite{Drabkin1969}.
We found for this sample $p$= 1.5$\times$10$^{21}$ cm$^{-3}$. The values of $p$ and $\gamma$ for the  samples under
consideration and the estimated  values of $p$ are listed in Table \ref{tab1}.

\begin{table*}
\caption{\label{tab1} Values of the hole concentration of our  samples
 in cm$^{-3}$ as determined by various methods (see text).
Also, values of the Sommerfeld coefficients $\gamma$ in mJ/molK$^2$ and $\beta$ = 2$\times${12}/{5}$\pi^4$ R/$\theta_{\rm Debye}^3$ in mJ/molK$^4$ are given as well as the value of $C_p/T^3$ at the maximum in  mJ/molK$^4$.
A blank signifies that a value is not available. The values from Ref. \onlinecite{Bevolo1976} are also cited for comparison.}
\begin{ruledtabular}
\begin{tabular}{ccccccc}
sample      & $p_{\rm H}$ & $\gamma$ & $\beta$ & $C_p/T_{\rm{max}}^3$  & $p_{\gamma}$ &$\gamma_{\rm op}$  \\ \hline
Bevolo  & 8$\times$10$^{19}$ & 0.554(3) & 0.307(1) &  0.640(2) & 8$\times$10$^{19}$ & - \\
CG1   & 1.6$\times$10$^{21}$ & 1.66(9) & 0.380(9) & 0.663(2) & 1.2$\times$10$^{21}$ & - \\
CG2  & - & 1.57(2) & 0.368(3) & 0.663(2) & 1 $\times$10$^{21}$  & 1.5 $\times$10$^{21}$ \\
CG3 & - & 1.19(2) & 0.339(2) & 0.658(2) & 4 $\times$10$^{20}$  & - \\
Alfa & - & 1.45(4) & 0.358(3) & 0.656(2) & 8 $\times$10$^{20}$  & - \\
\end{tabular}
\end{ruledtabular}
\end{table*}

Table \ref{tab1} displays the hole concentration  of our two samples and that given by
Bevolo~\cite{Bevolo1976} as determined by Hall measurements ($p_{\rm H}$) together with the concentration
estimated from the ir reflectivity ($\gamma_{\rm op}$) and those obtained from the measured values
of $\gamma$. The latter were estimated by scaling the values of
Bevolo \textit{et al.} using Eq. (\ref{eq1}).
In doing so, we have taken into account the fact that the effective mass $m^*$
increases with increasing $p$ due to nonparabolicity of the bands. Since $p_{\gamma}$ is proportional
to ($m^*$)$^3$, its estimated value depends very critically on this nonparabolic
increase, which is not accurately known at present. We have taken for this increase the
values reported in Ref. \onlinecite{Lewis1973}: 24\% for
the Alfa Aesar sample and 32\% for the CG sample with respect to Bevolo's result. In
spite of some variations, the values of $p$ listed in this table are adequate for
the subsequent discussion of our heat capacity measurements.

For the sake of completeness we also measured the Raman spectrum of the LO and TO
phonons in our CG-GeTe sample. We found at 293 K $\omega_{\rm LO}$ = 123 cm$^{-1}$ and  $\omega_{\rm TO}$ = 88 cm$^{-1}$,
in good agreement with the values reported
in Ref. \onlinecite{Steigmeier1970} for the same temperature.

\subsection{Temperature dependence of the heat capacity}

\begin{figure}[tbph]
\includegraphics[width=8.5cm ]{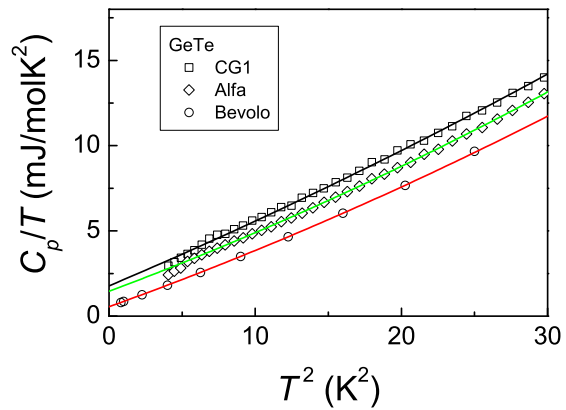}
\caption{(color online)
Sommerfeld plots of the heat capacities of our CG and Alfa samples of $\alpha$-GeTe together with that for Bevolo's sample.\cite{Bevolo1976}
The  $\gamma$
values obtained from the intercept with the vertical axis are listed in Table \ref{tab1}.} \label{fig3}
\end{figure}

We display in Fig. \ref{fig3} the heat capacity of GeTe measured below 5.5 K using the standard
Sommerfeld plot ($C_p/T$ vs. $T^2$) : The intercept of a linear fit with the ordinate yields the
value of $\gamma$ as listed in Table \ref{tab1}.

\begin{figure}[tbph]
\includegraphics[width=8.5cm ]{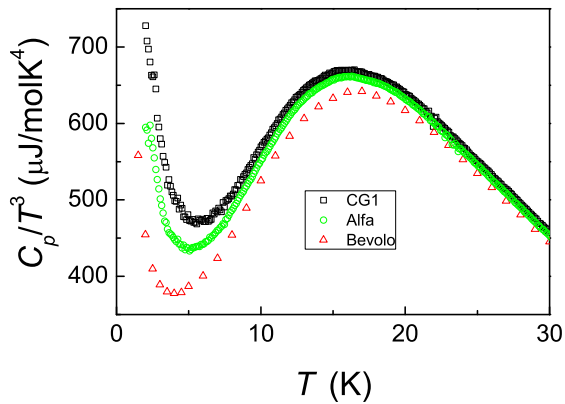}
\caption{(color online)
Measured temperature dependence of $C_p/T^3$  for the three samples of $\alpha$-GeTe  under consideration.
Details for other samples are given in Table \ref{tab1}.
Note the increase of $C_p/T^3$ with increasing
hole concentration (see Table \ref{tab1}).} \label{fig4}
\end{figure}

Figure \ref{fig4} shows the temperature dependence of $C_p/T^3$ measured for our two samples in
the 2.5 - 30K temperature range, together with the results of Bevolo \textit{et al.}.\cite{Bevolo1976} The increase
in $C_p/T^3$ seen below 5 K corresponds to the Sommerfeld term shown in Fig. \ref{fig3}. The most
remarkable feature of Fig. \ref{fig4} is the monotonic increase in $C_p/T^3$ observed with
increasing $p$, which can be assigned to the frequency softening (cf. Fig. \ref{fig1} and the
corresponding down-shift of the DOS (cf. Fig. \ref{fig2}(a)). Notice also that our two samples
exhibit this maximum at 15.8 K whereas Bevolo's maximum is found at 16.6 K. This
shift can be assigned to the fact that the latter sample has a hole concentration
over an order of magnitude lower than the former.

A simple relationship has been proposed earlier (Ref. \onlinecite{Cardona2007,Romero2008})
to relate the temperature of the maximum in $C_p/T^3$
to the phonon DOS. One first
divides the temperature by 1.43  so as to convert $T$ (in K) into cm$^{-1}$.
Then one divides the obtained frequencies by 6.2. This procedure leads to the
frequencies 68.5 cm$^{-1}$ for our samples and 72 cm$^{-1}$ for Bevolo's sample.\cite{Bevolo1976} The corresponding
shift of 3.5 cm$^{-1}$ agrees with that observed for the first TA peak in the DOS.
The position of this peak, 70 cm$^{-1}$  also agrees with the average position estimated
by the procedure  followed above.

\begin{figure}[tbph]
\begin{center}
\includegraphics[width=7.5cm,clip=true]{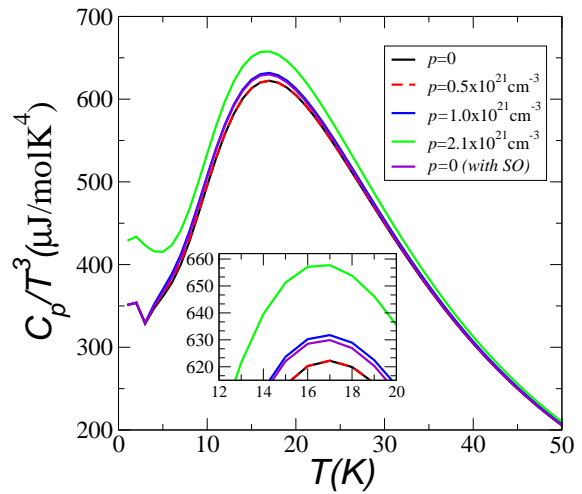}
\end{center}
\caption{(color online)
Calculated temperature dependence of $C_v/T^3$ for an undoped $\alpha$-GeTe sample (with and without SO splitting),
 and for a sample with $p=0.5\times 10^{21}$,
$1.0\times 10^{21}$, and $2.1\times10^{21}$ cm$^{-3}$ (without SO splitting).
} \label{fig5}
\end{figure}

Figure \ref{fig5} displays the temperature dependence of $C_v/T^3$ ($\sim C_p/T^3$ ) calculated
for several different concentrations ($p$=0, 0.5$\times$10$^{21}$, 1.0$\times$10$^{21}$,
and 2.1$\times$10$^{21}$ cm$^{-3}$). Notice that the effect of SO coupling, as implemented for the undoped sample, is small
 and we have not implemented it for doped GeTe. Hence, in the latter case we should compare
it with the undoped calculation without SO
interaction. We did not notice much of a change  for the light doping case ($p$=0, 0.5$\times$10$^{21}$ cm$^{-3}$).
However, as expected, the increase in doping concentration leads to a monotonic increase in the maximum
of $C_v/T^3$. As seen in Figure \ref{fig5}, we found an increase in the maximum of
0.2 and 0.6\% for $p$=1.0$\times10^{21}$ and 2.1$\times$ 10$^{21}$ holes/cm$^{3}$, respectively.
The results  agree rather well with the 0.5\% measured for the CG sample, especially when one
considers that  this sample has a doping of about 1.5$\times$10$^{21}$ cm$^{-3}$.
The shift in the calculated position of the maximum with doping is also close to the measured one.

\begin{figure}[tbph]
\includegraphics[width=8.5cm ]{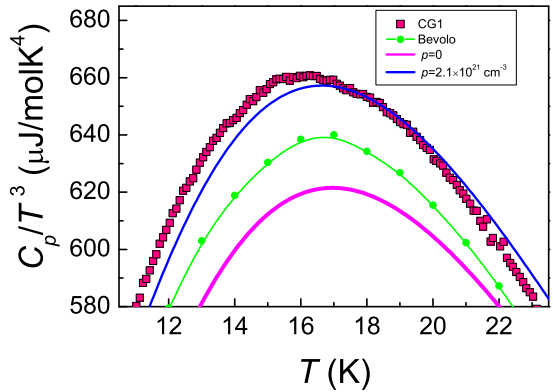}
\caption{(color online)
Heat capacity divided by $T^3$ measured for samples (Bevolo and our data CG1) compared with calculations with $p$=0 and $p$=2.1$\times$10$^{21}$ cm$^{-3}$
without SO coupling.} \label{fig6}
\end{figure}

It is of some interest to compare the small effect of the SO interaction shown in Fig. \ref{fig5}
with that calculated for PbTe. In Ref. \onlinecite{Romero2008}
we proposed a perturbation expansion of the SO effect as a function of the strength
of the SO splittings of the $p$ valence electrons of Pb and Te.
We use the same expression replacing the SO splitting of Pb (1.26 eV, cf. Ref. \onlinecite{Herman1963})
by that of Ge (0.22 eV, cf. Ref. \onlinecite{Herman1963}) and predict an effect of the SO coupling
of 4\% for the maximum of
$C_v/T^3$ in $\alpha$-GeTe (as opposed to 20\% for PbTe). The effect observed in Fig. \ref{fig5} is about 2\%,
somewhat smaller than the rough estimate carried over from the PbTe.

\begin{figure}[tbph]
\includegraphics[width=8.5cm ]{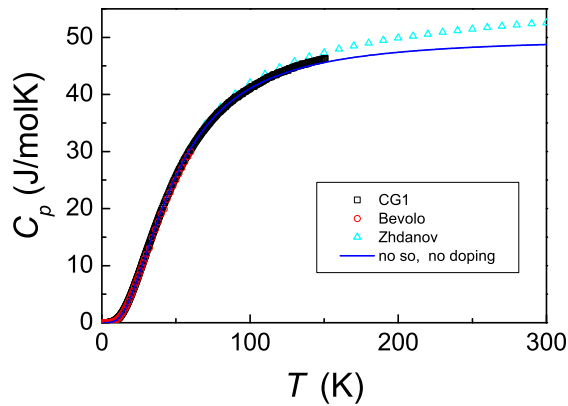}
\caption{(color online)
Measured heat capacity of several $\alpha$-GeTe samples in the 3-300 K
range. The black solid line represents our ab initio calculations. The linear dependence in Zhdanov's data for $T <$ 150 K corresponds to the thermal expansion effect.} \label{fig7}
\end{figure}

We conclude by displaying in Fig. \ref{fig6} the temperature dependence of $C_p$ as measured
by us, Bevolo \textit{et al.}\cite{Bevolo1976} and Zhdanov~\cite{Zhdanov1971}, compared with our \textit{ab initio} calculations
of  $C_v$ up to 300 K. Our measurements up to 150 K, from  which the Sommerfeld term has been subtracted, agree within error, with those of Zhdanov.
Between 150 and 300 K, Zhdanov's data lie above the calculations. In order to ascertain whether this discrepancy is due to the difference between $C_p$ and $C_v$ we use the expression:

\begin{equation}
C_p(T) - C_v(T) = \alpha_v^2(T) \cdot B \cdot V_{\rm{mol}} \cdot T,
\label{EqGaN}
\end{equation}\\

where $\alpha_v(T)$ is the temperature dependent  coefficient of the volume thermal expansion, $B$ is the bulk modulus and $V_{\rm mol}$ is the molar volume.

We have evaluated Eq.(\ref{EqGaN}) at 300 K using
$\alpha_v$=5.60$\times$10$^{-5}$ K$^{-1}$ (Ref. \onlinecite{Wiedemeyer1977}), $B$ = 49.9 GPa (Refs. \onlinecite{Onodera,Shaltaf2008}) and a molar volume of 32.3 cm$^3$. We find a thermal expansion contribution to $C_p$ - $C_v$ of $\sim$ 1.5 J/molK at 300 K, which is by a factor of $\sim$2.5 smaller than the value required to bring Zhdanov's data ($C_p$) to agree with the calculation ($C_v$). In order to clarify this matter, measurements should
be repeated in the temperature range 150 - 300 K.

\section{Conclusions}
We have presented a detailed experimental and theoretical investigation of
lattice dynamics and heat capacity of  $\alpha$-GeTe.
In contrast to previously reported results for other IV-VI materials~\cite{Romero2008},
we found that the inclusion of spin orbit coupling has a small effect
on the calculated phonon frequencies and consequently on the heat capacity.
On the other hand, we found that ignoring non-stoichiometry effects of GeTe,
leads to discrepancies between the calculated and the measured  $C_{p}/T^3$.
These can be resolved by taking into account the existence of
free holes which lead to an increase of $C_{p}/T^3$. Such increase was found to be due
to phonon frequency softening and the corresponding down shift of the density of phonon states.

\begin{acknowledgments}
We acknowledge financial support by the Interuniversity Attraction
Poles Program (P6/42) - Belgian State - Belgian Science Policy.
Two of the authors (R.S. and X.G.) acknowledge support from the
the Communaute Francaise de Belgique (Action de Recherches
Concertee 07/12-003) and the European Union (NMP4-CT-2004--500198,
``NANOQUANTA'' Network of Excellence ``Nanoscale Quantum
Simulations for Nanostructures and Advanced Materials'', and
``ETSF'' Integrated Infrastructure Initiative).  We thank E. Sch\"onherr for sample preparation.
We also thank  M. Giantomassi for providing
the subroutines for the calculations of atomic projected density of states.

\end{acknowledgments}

\end{document}